\documentstyle[twocolumn,seceq]{jpsj}
\def\geqap{\,\raise 2pt \hbox{$>\kern-11pt \lower 5pt \hbox{$\sim$}$}\,}
\def\leqap{\,\raise 2pt \hbox{$<\kern-10pt \lower 5pt \hbox{$\sim$}$}\,}
\def\runtitle{Optical Conductivity in NaV$_2$O$_5$}
\def\runauthor{Satoshi {\sc Nishimoto} and Yukinori {\sc Ohta}}
\title{Optical Conductivity of the Trellis-Lattice $t$-$J$ Model: 
\\ Charge Fluctuations in NaV$_2$O$_5$}
\author{Satoshi {\sc Nishimoto}$^{1}$ and Yukinori {\sc Ohta}$^{1,2}$}
\inst
{$^1$ Graduate School of Science and Technology, Chiba University, 
Inage-ku, Chiba 263-8522\\
$^2$Department of Physics, Chiba University, Inage-ku, Chiba 263-8522}
\recdate{August 6, 1998}
\abst
{
Optical conductivity of the trellis lattice $t$-$J$ model 
at quarter filling is calculated by an exact-diagonalization 
technique on small clusters, whereby the valence state of V 
ions of NaV$_2$O$_5$ is considered.  
We show that the experimental features at $\sim$1 eV, including 
peak positions, presence of shoulders, and anisotropic spectral 
weight, can be reproduced in reasonable range of parameter values, 
only by assuming that the system is in the charge disproportionated 
ground state.  
Possible reconciliation with experimental data suggesting the 
presence of uniform ladders at $T>T_c$ is discussed.  
}
\kword
{NaV$_2$O$_5$, optical conductivity, $t$-$J$ ladder, charge fluctuation, 
trellis lattice, spin-Peierls}
\begin{document}
\sloppy
\maketitle

The valence state of V ions in NaV$_2$O$_5$, a compound once believed 
to be a typical spin-Peierls system\cite{isobe,fujii}, has become an 
issue of intense controversy even above the transition temperature 
$T_c$;  recent X-ray structural analysis and other 
experiments\cite{xray1,xray2,xray3,nmr2} seem to suggest the uniform 
valence of V$^{4.5+}$, the system being the coupled uniform ladders at 
quarter filling, whereas an interpretation of low-energy optical 
conductivity data suggests the system to be in a broken-parity electronic 
ground state.\cite{opt2}  To clarify this issue will help one understand 
the mechanism of the phase transition of this compound observed at $T_c=34$ K.   

In this letter, we consider this issue by calculating the optical 
conductivity $\sigma(\omega)$ of the trellis-lattice $t$-$J$ model at 
quarter filling directly.  We will show that the experimental features 
of $\sigma(\omega)$ at $\omega\simeq 0.6-2.5$ eV, such as the positions 
of the main peak and its shoulders, anisotropy in the spectral weight, etc., 
observed\cite{opt2,opt1} over a wide temperature range above and below 
$T_c$, can be explained within a reasonable range of parameter values, 
only by assuming that the system is in the charge disproportionated 
ground state.  We will thereby point out that the difference in the time 
scale of the measurements may be a possible way of reconciliation: 
charge (or valence) fluctuation of V ions is slow compared with the 
frequency of electric field of light used in the optical measurement, 
but it is fast enough compared with the time scale of the other 
measurements.  We would suggest that measurements of frequency range 
of $\sim$50 GHz to $\sim$1 THz should be able to detect a possible 
resonant charge fluctuation in this compound.  
\begin{figure}
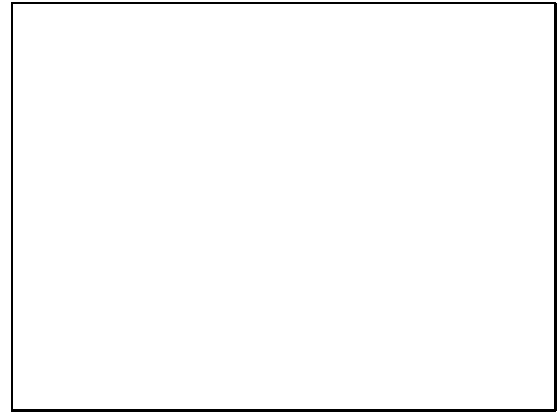

\figureheight{5.2cm}
\caption{
Schematic representation of the trellis lattice $t$-$J$ model.  
The structure consists of the anisotropic ladders (solid lines) 
connected with zigzag chain bonds (dotted lines).  
The crystallographic ${\mib a}$- and ${\mib b}$-axes are also shown.  
}
\label{fig:1}
\end{figure}

The low-energy electronic state of the two-dimensional plane of 
NaV$_2$O$_5$ may be written by the trellis-lattice $t$-$J$ model, 
i.e., coupled anisotropic $t$-$J$ ladders (see Fig.~1), as we have 
shown in ref.\cite{nishimoto}  The Hamiltonian is given by 
\begin{eqnarray}
H=&-&\sum_{\langle ij\rangle\sigma}t_{ij}
\big(\hat{c}^\dagger_{i\sigma}\hat{c}_{j\sigma}+{\rm H.c.}\big)
\nonumber \\
&+&\sum_{\langle ij\rangle}J_{ij}\Big({\mib S}_i\cdot{\mib S}_j
-\frac{1}{4}n_in_j\Big)
\nonumber \\
&+&\sum_{<ij>}V_{ij}n_in_j
+\sum_i\Delta_i n_i~,
\end{eqnarray}
where $\langle ij\rangle$ represents the nearest-neighbor 
bonds along the legs and rungs with the hopping and exchange parameters 
$t_{ij}$ and $J_{ij}$ taking $t_\perp$ and $J_\perp$ for the rungs 
and $t_\parallel$ and $J_\parallel$ for the legs.  It also represents 
the zigzag chain bonds with parameters $t_{xy}$ and $J_{xy}$ where 
the intersite Coulomb repulsion $V_{xy}$ is also taken into account.  
We introduce the `symmetry-breaking' on-site energy of an electron, 
$\Delta_i$, which takes either $\Delta$ ($\geq 0$) or 0 depending on 
sites, whereby we can make the system charge disproportionated if 
$\Delta\ne 0$.  
$\hat{c}^\dagger_{i\sigma}=c^\dagger_{i\sigma}(1-n_{i-\sigma})$ 
is the constrained electron-creation operator at site $i$ and 
spin $\sigma$ $(=\uparrow,\downarrow)$, ${\mib S}_i$ is the 
spin-$\frac{1}{2}$ operator, and $n_i=n_{i\uparrow}+n_{i\downarrow}$ 
is the electron-number operator.  
We consider the case at quarter filling.  

We employ a Lanczos exact-diagonalization technique on small clusters 
and calculate the frequency-dependent conductivity $\sigma(\omega)$ to 
examine the anisotropic optical response of the system.  A finite-size 
cluster of the trellis lattice, i.e., coupled two $4\times 2$ ladders 
shown in Fig.~1, is used with periodic boundary condition.  
The optical conductivity is defined by 
\begin{equation}
\sigma_\alpha(\omega)=\sum_{\nu\ne 0}\,{1\over\omega}\,
|\langle\psi_\nu |j_\alpha |\psi_0\rangle|^2
\,\delta(\omega-(E_\nu-E_0)), 
\end{equation}
where $|\psi_\nu\rangle$ $(E_\nu)$ denotes the $\nu$-th eigenstate 
(eigenenergy) of the system (in particular, $\nu=0$ denotes the 
ground state).  Also, $j_\alpha$ with $\alpha=a,\,b$ denotes a component 
of the current operator 
\begin{equation}
j_\alpha=i\sum_{ij}t_{ij}({\mib r}_i-{\mib r}_j)_\alpha
\hat{c}^\dagger_{i\sigma}\hat{c}_{j\sigma}, 
\end{equation}
where $({\mib r}_i-{\mib r}_j)_\alpha$ is the $\alpha$-component of the 
vector connecting between sites $i$ and $j$.  
The system we treat is in the insulating state\cite{nishimoto} and there 
is no Drude peak at $\omega=0$.  

We use the following two sets of values of the ladder parameters 
(in units of eV) 
obtained from the $d$-$p$ model by the method of ref.\cite{nishimoto}.  
Set A: 
$t_\perp=0.298$, 
$t_\parallel=0.140$, 
$J_\perp=0.0487$, and 
$J_\parallel=0.0293$, 
which are obtained in ref.\cite{nishimoto} using realistic values of the 
$d$-$p$ model parameters $t_{pd}=1.22$ ($1.03$) for the hopping integral 
of the rung (leg) V-O bonds, $U_d=4.0$ for the on-site Coulomb repulsion 
on V ions, and $\Delta_{pd}=4.0$ ($6.5$) for the energy-level difference 
between the V $d_{xy}$-orbital and O $p$-orbital on the rung (leg).  
Set B: 
$t_\perp=0.397$, 
$t_\parallel=0.194$, 
$J_\perp=0.0658$, and 
$J_\parallel=0.0546$, 
which are obtained using a different set of values of the $d$-$p$ model 
parameters with enhanced hopping amplitude: $t_{pd}=1.46$ ($1.24$), 
$U_d=5.0$, and $\Delta_{pd}=4.0$ ($6.5$) for the rung (leg) V-O bonds.  
Values of the other parameters $t_{xy}$, $V_{xy}$, and $\Delta$ are 
then varied for simulating various situations.  
We use the relation $J_{xy}=4t_{xy}^2/U_d$ for simplicity.  

The optical conductivity observed in experiment\cite{opt1,opt2} has 
the following features: 
(i) Onset of charge-transfer excitations from O $p_x$- or $p_y$- to 
V $d_{xy}$-orbitals appears at $\omega\simeq 3-3.5$ eV which is 
anisotropic between ${\mib a}\parallel{\mib E}$ and ${\mib b}\parallel{\mib E}$.  
(ii) There appear broad spectral features around $\omega\simeq 1$ eV, 
which have been identified as the $d$-$d$ transitions;\cite{opt1,opt2} 
the main peak has a shoulder in $\sim$0.5 eV ($\sim$0.8 eV) higher-energy 
side in $\sigma_a(\omega)$ ($\sigma_b(\omega)$).  
Also the height of the peaks (or the integrated spectral weight) is 
markedly larger in ${\mib a}\parallel{\mib E}$ than 
in ${\mib b}\parallel{\mib E}$.  
(iii) There is a small continuum in the low-energy region of 
$\omega\simeq 0.01-0.1$ eV, the origin of which was argued to be 
`charged magnons'.\cite{opt2}  The spectral features (i) and (ii) are hardly 
changed over a wide temperature range above and below $T_c$,\cite{popova} 
while the oscillator strength of the feature (iii) decreases suddenly 
at $T_c$ with lowering temperatures.  
We now want to discuss the origins of the features (ii).  
\begin{figure}
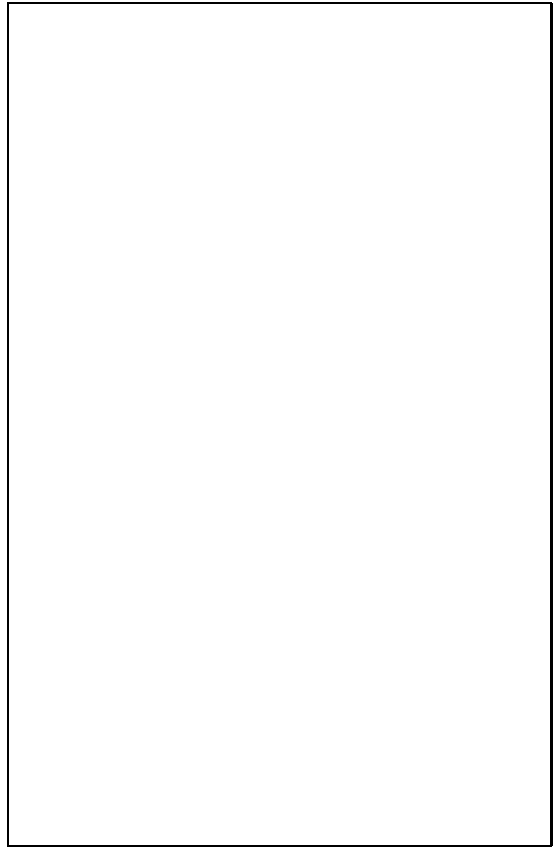

\figureheight{11.0cm}
\caption{
Optical conductivity $\sigma_\alpha(\omega)$ calculated for 
the trellis-lattice $t$-$J$ model with $\Delta=0$.  
Solid and dashed curves show $\sigma_a(\omega)$ for 
${\mib a}\parallel{\mib E}$ and $\sigma_b(\omega)$ for 
${\mib b}\parallel{\mib E}$, respectively.  
In (a) and (c), we use values of the ladder parameters of Set A, 
and in (b), we use those of Set B.  
All the $\delta$-functions in Eq.~(2) are Lorentzian 
broadened with the width of $0.05$ eV.  
Peaks at $\omega=0$, which appear due to the broadening 
and factor $1/\omega$ in Eq.~(2), are spurious.  
}
\label{fig:2}
\end{figure}

Let us first take some simplified pictures.  
Suppose an isolated rung of the ladder, then we have the $d$-$d$ 
excitation peak in $\sigma_a(\omega)$ at $\omega=\sqrt{(2t_\perp)^2+\Delta^2}$, 
the value of which is however too small compared to $\omega\simeq 1$ eV 
if $\Delta=0$ or there is no charge disproportionation,\cite{opt2} because the 
effective hopping amplitude $t_\perp\simeq 0.3$ eV strongly reduced from the 
bare $d$-$p$ hybridization $t_{pd}$ is relevant here.\cite{nishimoto}  
Suppose next the trellis lattice in the limit of small hopping amplitude 
and large $V_{xy}$ and/or $\Delta$ where the sites of, say, the right legs of the 
ladders are all with $\Delta>0$ and those of the left with $\Delta=0$ (resulting 
in the charge disproportionated ground state), then we have a peak in 
$\sigma_a(\omega)$ at $\omega=\Delta+2V_{xy}$.  This is because the electron 
transferred through the rung interacts (by $V_{xy}$) with two electrons on the 
leg of the neighboring ladder across the zigzag chain bonds; the peak position 
can be comparable to $\omega\simeq 1$ eV because $V_{xy}\simeq 0.5$ eV is enough 
even at $\Delta=0$.  If small randomness occurs in the disproportionated charge 
configuration, then another peak can also appear at $\omega=\Delta+V_{xy}$, 
because the transferred electron may interact with only one electron on 
the leg of the neighboring ladder across the zigzag chain bonds.  Thus, there 
appears a two-peak structure in $\sigma_a(\omega)$ (which, as we will see below, 
corresponds to the experimental broad features around $\omega\simeq 1$ eV).  
It is also noted that, if the charge configuration completely disproportionated 
on one side of the legs is assumed, the spectral weight for 
${\mib b}\parallel{\mib E}$ becomes quite small, because the electrons are hard to 
oscillate along the ${\mib b}$-direction due to the strong on-site Coulomb repulsion 
(or prohibited double occupancy in our model) and small values of $t_{xy}$; 
any randomness in this charge configuration would however make the spectral 
weight $\sigma_b(\omega)$ larger.  

Now let us see the calculated results for $\sigma_\alpha(\omega)$ with 
keeping these ideas in mind.  
We first of all find in Fig.~2(a), where we use the parameter Set A 
with $\Delta=0$ and $V_{xy}=0$ (i.e., without charge disproportionation), 
that the main features appeared at $\omega\simeq 0.5$ eV are again too low 
in energy compared with experiment.  The parameter Set B with $\Delta=0$ 
and $V_{xy}=0$ also gives similar results.  We also find that the 
relative main-peak positions between ${\mib a}\parallel{\mib E}$ and 
${\mib b}\parallel{\mib E}$ are sensitive to the changes in small parameter 
values $t_{xy}$ and $V_{xy}$ and that some broadened features in 
$\sigma_a(\omega)$ are induced by the presence of $t_{xy}$.  We however 
find that the essential spectral features, i.e., the overall main-peak 
positions, remain unchanged.  These results thus demonstrate that the 
$\omega\simeq 1$ eV features cannot be reproduced in reasonable parameter 
range, without charge disproportionation.  

The charge disproportionated ground state can be achieved by including 
either the large inter-site Coulomb repulsion $V_{xy}$ along 
the zigzag chain bonds\cite{nishimoto} or the on-site energy-level 
difference $\Delta$ between, say, the left and right legs of the ladders.  
In the former case, the calculated charge correlation functions 
$\langle n_in_j\rangle$ indicate that at $V_{xy}\geqap 0.6-0.7$ eV the 
strong tendency toward charge disproportionation comes out\cite{nishimoto} 
although the symmetry is not really broken in finite-size systems.  
Figures 2(b) and (c) show the results for such situations.  
Here we assume $V_{xy}=1.0$ eV, and use values of the parameter Set B 
with $t_{xy}=0.1$ eV and Set A with $t_{xy}=0.05$ eV in Figs.~2(b) and (c), 
respectively.  We find in both figures that 
the broad spectral features in $\sigma_a(\omega)$ around $\omega\simeq 1$ 
eV really appear and some smaller features also appear at $\omega\simeq 2$ 
eV.  We analyze the final states corresponding to the largest peaks and 
find that the features at $\sim$1 eV come mainly from the states with 
`$\omega\simeq V_{xy}$' and the features at $\sim$2 eV come mainly from 
the states with `$\omega\simeq 2V_{xy}$', as we have discussed above using 
simplified pictures.  The anisotropy in $\sigma_\alpha(\omega)$ between 
${\mib a}\parallel{\mib E}$ and ${\mib b}\parallel{\mib E}$ is also noted; 
the main peak in $\sigma_b(\omega)$ is located in $\sim$0.5 eV higher 
energy side than that in $\sigma_a(\omega)$ and the spectral weight for 
$\sigma_b(\omega)$ is considerably smaller than that for $\sigma_a(\omega)$.  
These results are all in good agreement with experiment.  
To get more precise agreement with experimental features would necessitates 
the calculations in larger size clusters, which are however not feasible 
in cluster diagonalizations.  
\begin{figure}
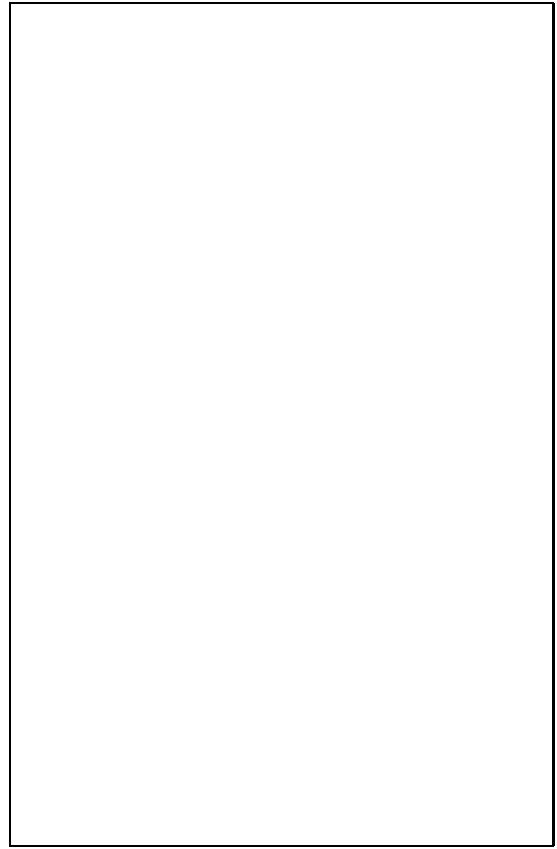

\figureheight{11.0cm}
\caption{
Same as Fig.~2 but for the case with $\Delta\ne 0$.  Here we assume that 
the sites on the left legs of the ladders are all with $\Delta>0$ and 
those on the right are all with $\Delta=0$.  Values of the ladder parameters 
of Set A are used.  
}
\label{fig:3}
\end{figure}

Next let us see the results obtained by introducing $\Delta\ne 0$ to 
break the symmetry and actually realizing the charge-disproportionated 
ground state.  
Note that although such an {\it ad hoc} assumption of the presence of 
nonzero $\Delta$ induces charge disproportionation in the system, 
its origin or the mechanism of appearing $\Delta\ne 0$ should be 
clarified; the simplest explanation\cite{nishimoto} may be the presence 
of strong intersite Coulomb repulsions, which induces the broken-symmetry 
ground state in the thermodynamic limit and leads to $\Delta\ne 0$.  
The calculated results for $\sigma_\alpha(\omega)$ are shown in Fig.~3 
where we assume that the sites of the right legs of the ladders are all 
with $\Delta>0$ and those of the left are all with $\Delta=0$.  
The ladder parameters of Set A are used.  
We assume $\Delta=0.3$, $V_{xy}=0.2$, and $t_{xy}=0.03$ eV in Fig.~3(a), 
which result in the valences of V$^{4.789+}$ and V$^{4.211+}$; here we 
find that the peak positions in $\sigma_\alpha(\omega)$ are again still 
too low compared with experiment.   
By increasing the values of $\Delta$ and $V_{xy}$, we find that 
the peak positions shift to higher energy side and the agreement 
with experiment becomes better; in Fig.~3(b), we assume $\Delta=0.4$, 
$V_{xy}=0.5$, and $t_{xy}=0.05$ eV, which result in the valences of 
V$^{4.940+}$ and V$^{4.060+}$, and in Fig.~3(c), we assume $\Delta=0.4$, 
$V_{xy}=0.5$, and $t_{xy}=0.03$ eV, which result in the valences of 
V$^{4.943+}$ and V$^{4.057+}$.  
In the latter two cases, we find that (i) the main peak appears at 
$\omega\simeq 1$ eV and its shoulders at $\omega\simeq 1.6$ eV 
in both $\sigma_a(\omega)$ and $\sigma_b(\omega)$ where the distance 
between the main peaks and shoulders is controlled primarily by the 
value of $V_{xy}$, (ii) the main-peak position in $\sigma_b(\omega)$ is 
slightly higher in energy than that in $\sigma_a(\omega)$, and (iii) the 
strong anisotropy of the spectral weight between ${\mib a}\parallel{\mib E}$ 
and ${\mib b}\parallel{\mib E}$ appears.  These results are all in good 
qualitative agreement with experiment.  
We furthermore find that the broad main peak has some internal 
structures, which may correspond to the experimentally observed 
broad peak which also has some structures.\cite{popova}  

We note that the anisotropy in the spectral weight between 
${\mib a}\parallel{\mib E}$ and ${\mib b}\parallel{\mib E}$ which is seen 
in Figs.~3(b) and (c) is too strong compared with the experimental anisotropy.  
This might suggest that the charge-ordering pattern assumed above 
should be made more `random', so that the electrons may oscillate along 
the legs of the ladders.  
To check this idea and to improve the strength of the anisotropy, we also 
use the charge-ordering pattern called the zigzag structure\cite{seo,khoms} 
which was argued to be realistic at $T<T_c$.  
The calculated result for $\sigma_\alpha(\omega)$ is shown in Fig.~4, 
where we indeed find that the spectral weight of $\sigma_b(\omega)$ 
is much increased without changing other features, resulting in a very good 
agreement with the experimental features.  

We have also sought for low-energy ($\omega\leqap 0.1$ eV) features 
argued as `charged magnons'\cite{opt2} in our calculated results.  
We however could not find any indications: there are many low-energy 
eigenstates but none of them has nonzero spectral weight.  
Interestingly, we find that, if we introduce a local lattice displacement, 
a small weight appears at the low energies, which might suggest the 
relevance of phonon-assisted processes.\cite{opt2}  
\begin{figure}
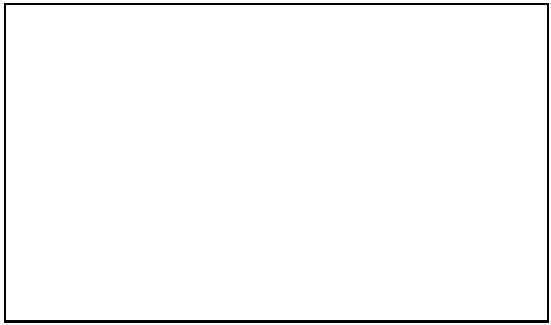

\figureheight{4.0cm}
\caption{
Same as Fig.~2 but for the case with $\Delta\ne 0$.  Here we assume 
that the sites corresponding to V$^{5+}$ ions on the zigzag 
structure of ordered electrons\cite{seo,khoms} have a nonzero value 
of $\Delta$.  Values of the ladder parameters of Set A are used.  
}
\label{fig:4}
\end{figure}

Finally, let us discuss the controversy in the valence state of 
V ions in NaV$_2$O$_5$ at $T>T_c$: recent X-ray structural 
analyses\cite{xray1,xray2,xray3} and NMR measurements\cite{nmr2} 
seem to suggest the uniform valence of V$^{4.5+}$, the system 
being the coupled uniform ladders with no charge 
disproportionation, which apparently contradicts our results 
shown above.  
However, we immediately notice that, if the dynamical fluctuation 
of the valence state of V ions is assumed, a possible way of 
reconciliation is to take into account the difference in the time 
scale of the measurements: 
if the charge (or valence) fluctuation of V ions is slow compared 
with the frequency of electric field of light used in the optical 
measurements, one finds the system to be charge-disproportionated, 
but if the charge fluctuation is fast enough compared with the 
time scales of the NMR measurement (an order of $1$$-$$100$ MHz) and 
other structural analysis experiments, one finds the system 
to be uniform.  
It is reported that an ESR experiment of up to 24 GHz finds no 
motional narrowing of the linewidth\cite{esr} and that the optical 
measurement down to the lowest frequency of $\sim$10 cm$^{-1}$ 
detects no anomalous behaviors indicating resonances.\cite{opt2,opt1}  
We would thus suggest that measurements of frequency range of 
$\sim$50 GHz to $\sim$1 THz should be able to detect the resonant 
charge fluctuation in this material.  
Precise characterization of the charge fluctuations at $T>T_c$ needs to 
be made further from both experimental and theoretical sides to 
clarify the mechanism of the phase transition of NaV$_2$O$_5$ at 
$T=34$ K.  

In summary, we have calculated the optical conductivity of the 
trellis lattice $t$-$J$ model at quarter filling using an 
exact-diagonalization technique on small clusters and considered 
the valence state of V ions of NaV$_2$O$_5$.  We have shown that 
the experimental features observed at $\omega\simeq 0.6-2.5$ eV, 
including the positions of the main peak and its shoulders, anisotropy 
in the spectral weight, etc., can be reproduced in reasonable range of 
parameter values, only by assuming that the system is in the charge 
disproportionated ground state.  
We have discussed the possible reconciliation with experimental data 
suggesting the presence of uniform ladders.  

We would like to thank T. Ohama and I. Yamada for enlightening 
discussions and M. N. Popova and A. B. Sushkov for informing us of 
the experimental data prior to publication.  
Financial support for S.~N. by Sasakawa Scientific Research 
Grant from the Japan Science Society and for Y.~O. by Iketani Science 
and Technology Foundation are gratefully acknowledged.  
Computations were carried out at Computer Centers of the Institute 
for Solid State Physics, University of Tokyo and the Institute for 
Molecular Science, Okazaki National Research Organization.

\end{document}